\documentclass[print]{revtex4}
\usepackage{graphicx}

\begin{document}
\title{Two-qubit Quantum Logic Gate in Molecular Magnets}
\author{Jing-Min Hou$^1$\footnote{Electronic address: jmhou@mail.nankai.edu.cn},Li-Jun Tian$^2$, and Mo-Lin Ge$^1$}
 \affiliation{$^1$Theoretical
Physics Division, Nankai Institute of Mathematics, Nankai
University, Tianjin, 300071, China\\ and Liuhui Center for Applied
Mathematics,
 Tianjin, 300071, China\\
$^2$Department of Physics, Shanghai University, Shanghai, 200436,
China}
\begin{abstract}
We proposed a scheme to realize a controlled-NOT quantum logic
gate in a dimer of exchange coupled single-molecule magnets,
$[\textrm{Mn}_4]_2$. We chosen the ground state and the three
low-lying excited states of a dimer in a finite longitudinal
magnetic field as the quantum computing basis and introduced a
pulsed transverse magnetic field with a special frequency. The
pulsed transverse magnetic field induces the transitions between
the quantum computing basis so as to realize a  controlled-NOT
quantum logic gate. The transition rates between the quantum
computing basis  and between the quantum computing basis and other
excited states are evaluated and analyzed. \\
PACS number(s): 03.67.Lx, 75.50.Xx, 75.10.Jm

\end{abstract}
\maketitle  In recent years, quantum computation remarkably
attracts the interests of the theoretical and experimental
physicists because of its extensive  prospects for the practical
applications. With the development of the integration techniques,
computational devices get smaller, eventually, the physical
principles of quantum mechanics must be taken into account. The
quantum algorithms\cite{grover,shor} discovered show that quantum
computation is more effective than classical one.  Many systems
are investigated theoretically and experimentally to realize
quantum computing, such as trapped ions\cite{cirac}, solid state
NMR\cite{gershenfeld,cory}, quantum dots\cite{barenco2,loss},
SQUID\cite{averin,makhin} and spin clusters\cite{loss1,loss2}.
Recently, quantum computation in molecular magnets is addressed.
Leuenberger and Loss proposed a scheme to realize Grover's
algorithms in molecular magnets such as $\textrm{Fe}_8$ and
$\textrm{Mn}_{12}$\cite{leuenberger}.

 Any quantum logic gate can be decomposed into  one-qubit
 rotation and two-qubit controlled-NOT gate operation\cite{barenco},
so the key point for quantum computing is how to realize an
arbitrary single-qubit operation gate and a two-qubit
controlled-NOT gate(or controlled-phase  gate), which are
assembled together to realize a universal quantum computing.
Realizing a two-qubit quantum logic gate in molecular magnets
requires that there are interactions between different molecular
particles. But for most molecular magnets such as $\textrm{Fe}_8$
and $\textrm{Mn}_{12}$, the interactions between molecules are
relatively weak and have been neglected in most studies.
Fortunately, the recent study of a dimerized single-molecule
magnets, $[\textrm{Mn}_4]_2$, showed that intermolecular exchange
interactions are not always negligible\cite{wernsdorfer}. This
system is
$[\textrm{Mn}_4\textrm{O}_3\textrm{Cl}_4(\textrm{O}_2\textrm{CEt})_3(\textrm{py})_3]_2$(called
$[\textrm{Mn}_4]_2$), a member of the
$[\textrm{Mn}_4\textrm{O}_3\textrm{Cl}_4(\textrm{O}_2\textrm{C}\textbf{R})_3(\textrm{py})_3]_2$
family, with $\textbf{R}=\textrm{Et}$. The supramolecular linkage
within $[\textrm{Mn}_4]_2$ introduces exchange interactions
between the $\textrm{Mn}_4$ molecules via the six C--H$\cdots$Cl
pathways and the Cl$ \cdots $Cl approach, which lead to noticeable
antiferromagnetic coupling between the $\textrm{Mn}_4$ units. Each
$\textrm{Mn}_4$ can be modelled as a `giant spin' of $S=9/2$ with
Ising-like anisotropy.

In this paper,we proposed a scheme to realize a controlled-NOT
quantum logic gate in a dimer of exchange coupled single-molecule
magnets. First, we don't consider the transverse exchange
interactions between two $\textrm{Mn}_4$ units, and choose the
ground state and the three low-lying excited states as quantum
computing basis. Then, we introduce  an oscillating transverse
magnetic field, which can induce transitions between the quantum
computing basis so as to realize a controlled-NOT gate in
molecular magnets. In our scheme, the two dipole-allowed
transitions are at different resonance frequencies, while the
oscillating magnetic field is in resonance with only one of them.
Finally, we evaluate the amplitudes of transitions between quantum
computing basis and the decaying rates of them due to the
transverse exchange interactions.

{\it The model}-- The corresponding Hamiltonian of each
$\textrm{Mn}_4$ unit is given by\cite{wernsdorfer}
\begin{equation}
H_i=DS_{zi}^2+g\mu_BB_zS_{zi},
\label{hi}
\end{equation}
where $i=1$ or $2$(referring to the two $\textrm{Mn}_4$ units of
the dimer) , $D$ is  the axial anisotropy constant, $\mu_B$ is the
Bohr magneton, $S_z$ is the easy-axis spin operator, $g$ is the
electronic g-factor, and $B_z$ is the applied longitudinal field.
The last  term in Eq.(\ref{hi}) is the Zeeman energy associated
with an applied field. The $\textrm{Mn}_4$ units within
$[\textrm{Mn}_4]_2$  are coupled by a weak superexchange via both
the six C--H$\cdots$Cl pathways and the Cl$ \cdots $Cl approach.
Thus, the Hamiltonian of a dimer of  exchange coupled molecular
magnets, $[\textrm{Mn}_4]_2$, can be written as\cite{wernsdorfer}
\begin{eqnarray}
H_{dimer}=H_1+H_2+J_zS_{z1}S_{z2}+\frac{J_{xy}}{2}(S^+_{1}S^-_{2}+S^-_{1}S^+_{2}),
\label{dimer}
\end{eqnarray}
where  $J_z$ and $J_{xy}$ are respectively the longitudinal and
the transverse superexchange interaction constants, $S^+_{i}$ and
$S^-_{i}$ are the usual spin raising and lowering operators
respectively.

Firstly we  don't consider the transverse exchange interaction
term in Eq.(\ref{dimer}). Then the corresponding Hamiltonian can
be simplified to the form
\begin{equation}
H_0=H_1+H_2+J_zS_{z1}S_{z2}. \label{h0}
\end{equation}
For this simplification,  every eigenstate of $[\textrm{Mn}_4]_2$
can be labelled by two quantum numbers $(m_1,m_2)$ for two
$\textrm{Mn}_4$ units, with $m_1=9/2,7/2,\cdots,-9/2$ and
$m_2=9/2,7/2,\cdots,-9/2$. The corresponding eigenvalues are given
by
\begin{eqnarray}
E^{(0)}_{m_1m_2}=(m_1^2+m_2^2)D+(m_1+m_2)g\mu_BB_z+m_1m_2J_z,
\label{emn}.
\end{eqnarray}

To realize a controlled-NOT gate, we introduce a pulsed
time-depending transverse magnetic field
$V(t)=B_\bot(t)[\cos(\omega t)\textbf{e}_x-\sin(\omega
t)\textbf{e}_y]$, where $\textbf{e}_x$ and $\textbf{e}_y$ are
respectively the unit vectors pointing along the $x$ and $y$ axes.
Thus, we obtain the Hamiltonian due to the interactions with the
transverse magnetic field as\cite{berman}
\begin{eqnarray}
H_\bot(t)&=&-\sum_ig\mu_BB_\bot(t)[\cos(\omega
t)S_{xi}-\sin(\omega t)S_{yi}]\nonumber\\
&=&-\sum_i\frac{g\mu_BB_\bot(t)}{2}[e^{i\omega
t}S^+_{i}+e^{-i\omega t}S^-_{i}]. \label{ht}
\end{eqnarray}
The pulsed transverse magnetic field  rotates clockwise and thus
produces left circularly polarized $\sigma^-$ photons. Absorption
(emission) of $\sigma^-$ photons give rise to $\Delta m=-1(\Delta
m=+1)$ transitions of  spin states.

\begin{figure}[ht]
 \includegraphics[width=1.0\columnwidth]{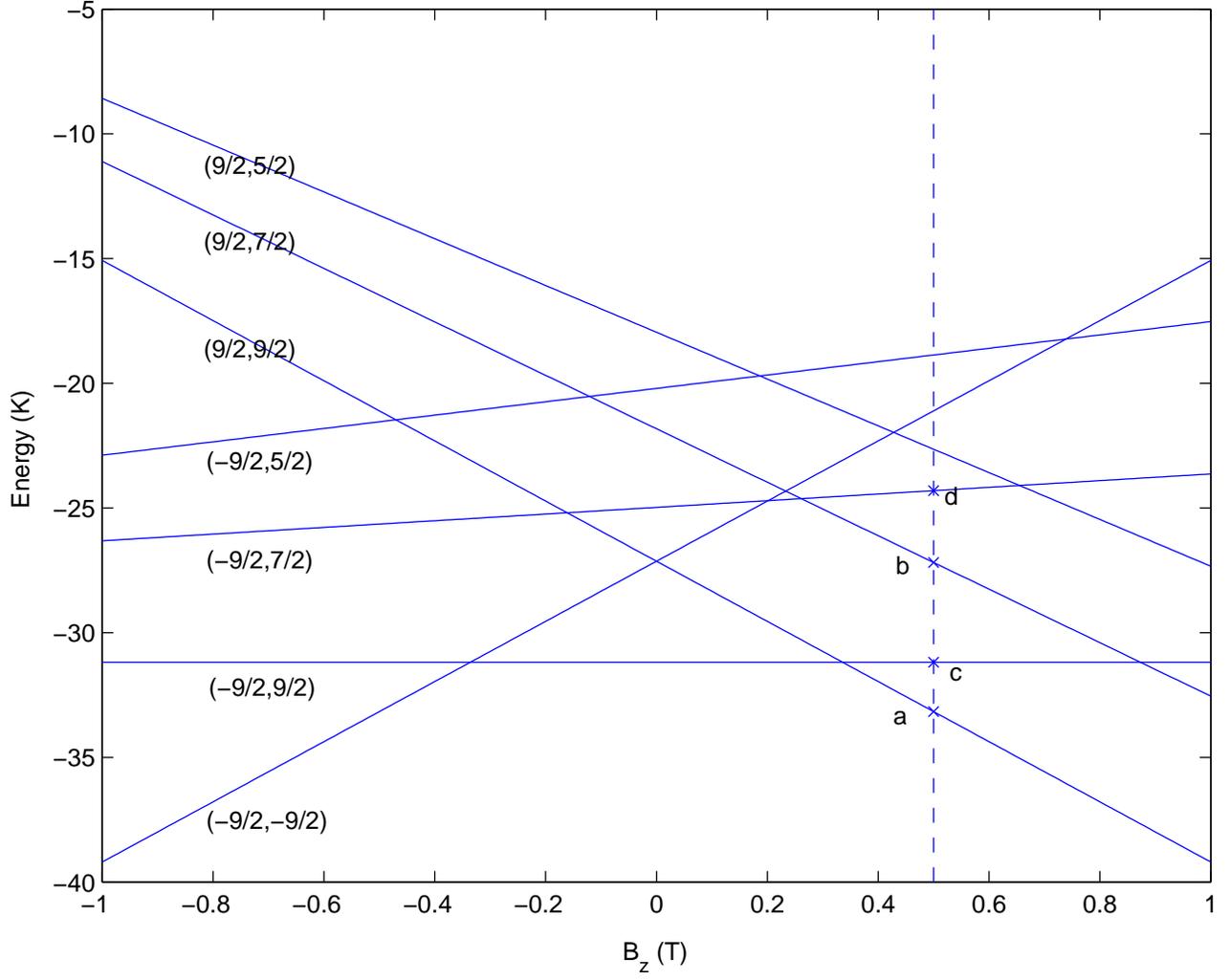}
 \caption{The spin states energies of $[\textrm{Mn}_4]_2$ for the low-lying states as
 a function of applied longitudinal magnetic field. The diagram is drawn according to the data
 calculated when $D=-0.72K$ and $J_z=0.1K$. Here, the states marked by
 the symbol $\times$ and labelled by
 the letters `a, b, c' and `d' are  chosen as
 the quantum computing basis in our scheme.}\label{1}
\end{figure}

{\it A controlled-NOT gate}--  We choose the ground state and the
three low-lying excited states in a finite longitudinal magnetic
field as the basis for quantum computing, which are marked by the
symbol `$\times$' and labelled respectively by the letters `a,b,c'
and `d' in Figure \ref{1}. These states are
$(9/2,9/2),(9/2,7/2),(-9/2,9/2),(-9/2,7/2)$ in a 0.5T longitudinal
magnetic field.

From the Eq.(\ref{emn}), we obtain the energy gaps between the
quantum computing  basis  or between them and neighboring excited
states, which are shown in Table \ref{tab1}. The energy gap
$E^{(0)}_{-9/2,7/2}-E^{(0)}_{-9/2,9/2}$ between the states
(-9/2,9/2) and (-9/2,7/2) is different from others, which is
important to realize a controlled-NOT gate in our scheme.
\begin{table}[ht]
\begin{center}
\caption{The energy gaps between the states chosen for quantum
computing  basis  or between them and the neighboring excited
states } \label{tab1}
\begin{tabular}{lc}\hline\hline
Energy gaps between states&\hspace{2mm} Values of  energy gaps\\
\hline
$E^{(0)}_{-9/2,7/2}-E^{(0)}_{-9/2,9/2}$&$-8D-g\mu_BB_z+\frac{9}{2}J_z$\\
$E^{(0)}_{9/2,7/2}-E^{(0)}_{9/2,9/2}$&$-8D-g\mu_BB_z-\frac{9}{2}J_z$\\
$E^{(0)}_{-9/2,5/2}-E^{(0)}_{-9/2,7/2}$&$-6D-g\mu_BB_z+\frac{9}{2}J_z$\\
$E^{(0)}_{9/2,5/2}-E^{(0)}_{9/2,7/2}$&$-6D-g\mu_BB_z-\frac{9}{2}J_z$\\
$E^{(0)}_{7/2,9/2}-E^{(0)}_{9/2,9/2}$&$-8D-g\mu_BB_z-\frac{9}{2}J_z$\\
$E^{(0)}_{7/2,7/2}-E^{(0)}_{9/2,7/2}$&$-8D-g\mu_BB_z-\frac{7}{2}J_z$\\
$E^{(0)}_{7/2,9/2}-E^{(0)}_{9/2,7/2}$&$0$\\
$E^{(0)}_{-7/2,7/2}-E^{(0)}_{-9/2,9/2}$&$-16D+8J_z$\\
$E^{(0)}_{-7/2,5/2}-E^{(0)}_{-9/2,7/2}$&$-14D+7J_z$\\

\hline\hline
\end{tabular}
\end{center}
\end{table}

\begin{figure}[ht]
 \includegraphics[width=0.9\columnwidth]{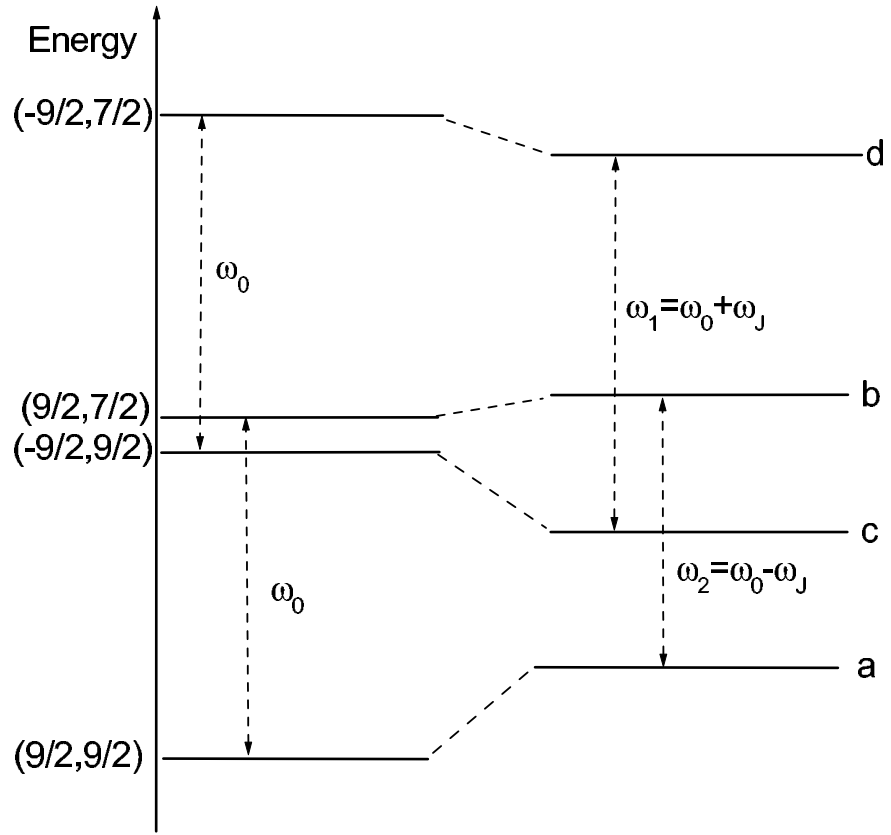}
 \caption{The schematic diagram for energy levels of a dimer $[\textrm{Mn}_4]_2$ without(the left hand) and
 with(the right hand) exchange interactions  between two $\textrm{Mn}_4$
 units. In the diagram, `a, b, c' and `d' refer to the states in Figure \ref{1}
 labelled by `a, b, c' and `d' respectively, which are the quantum
 computing basis in our scheme. Here $\omega_0$ and $\omega_1,\omega_2$ are the energy gaps between the quantum
 computing bases without and with exchange interactions respectively,  and $\omega_J=9J_z/2\hbar$ refers to
 the energy shift due to exchange interactions.}\label{2}
\end{figure}

Now we consider the pulsed transverse magnetic field  introduced
and evaluated the transition rates between the states by
considering the Hamiltonian term about the transverse magnetic
field, \textit{i.e.} Eq.(\ref{ht}), as a perturbation. Using a
rectangular pulse shapes with $B_\bot(t)=B_\bot$, if $-T/2<t<T/2$,
and $0$ otherwise, we obtain the quantum amplitude for the
transition from the state $(l,l')$ to $(k,k')$  induced by the
magnetic field pulse,
\begin{eqnarray}
C_{kk';ll'}&=&\frac{\pi g\mu_BB_\bot}{i\hbar}\sum_i \left[\langle
kk'|S^+_{i}|ll'\rangle\delta^{(T)}(\omega_{kk',ll'} +\omega
)+\langle
kk'|S^-_{i}|ll'\rangle\delta^{(T)}(\omega_{kk',ll'}-\omega
)\right]
\end{eqnarray}
where $\delta^{(T)}(\omega)={1}/{2\pi}\int_{-T/2}^{+T/2}e^{i\omega
t}dt={\sin(\omega T/2)}/{\pi\omega} $ is the delta-function of the
width $1/T$, ensuring overall energy conservation for $\omega T\gg
1$. For convenience, we denote the energy gaps between the quantum
computing basis as
$\omega_1=(E^{(0)}_{-9/2,7/2}-E^{(0)}_{-9/2,9/2})/\hbar$ and
$\omega_2=(E^{(0)}_{9/2,7/2}-E^{(0)}_{9/2,9/2})/\hbar$. From the
Table \ref{tab1}, we obtain $\Delta
\omega\equiv\omega_1-\omega_2=9J_z/\hbar$. In our scheme, we
choose the frequency of the magnetic pulse $\omega=\omega_1$.
 Then,the transition rate $w_{-9/2,7/2;-9/2,9/2}=|C_{-9/2,7/2;-9/2,9/2}|^2/T$ from the
 states
  $(-9/2,9/2)$ to $(-9/2,7/2)$ by absorbing a $\sigma^-$ photon  is
\begin{equation}
w_{-9/2,7/2;-9/2,9/2}\simeq\frac{9T(g\mu_BB_\bot)^2}{4\hbar^2},
\label{tran}
\end{equation}
where the relations $|\delta^{(T)}(\omega)|^2\approx
(T/2\pi)\delta^{(T)}(\omega)$ and $\delta^{(T)}(0)=T/2\pi$ are
used. The transition  rate $w_{-9/2,9/2;-9/2,7/2}$ from the states
$(-9/2,7/2)$ to $(-9/2,9/2)$ by emitting a $\sigma^-$ photon is
identical to $w_{-9/2,7/2;-9/2,9/2}$. Since
 the magnetic pulse frequency $\omega$ isn't equal to the energy
gap $\omega_2$ between the states (9/2,9/2) and (9/2,7/2),
\textit{i.e.}, $\omega_1\neq\omega_2$, which is shown in Figure
\ref{2}, the transition rate between the two states is very small
and can be negligible. From Reference \cite{wernsdorfer},
 the parameters $D$ and
$J_{z}$ are chosen as $ -0.72K$ and $0.1K$ respectively in this
paper. We insert the parameters $T=10^{-8}\textrm{s}$ and
$B_\bot=3.8\textrm{G}$ into Eq.(\ref{tran}) giving the transition
rates $w_{-9/2,7/2;-9/2,9/2}=w_{-9/2,9/2;-9/2,7/2}=1.0\times 10^8
\textrm{s}^{-1}$, while
$w_{9/2,7/2;9/2,9/2}=w_{9/2,9/2;9/2,7/2}=2.4\times 10^2
\textrm{s}^{-1}$. If we set $T=10^{-7}\textrm{s}$ and
$B_\bot=0.38\textrm{G}$, then the transition rates
$w_{-9/2,7/2;-9/2,9/2}=w_{-9/2,9/2;-9/2,7/2}=1.0\times 10^7
\textrm{s}^{-1}$, while
$w_{9/2,7/2;9/2,9/2}=w_{9/2,9/2;9/2,7/2}=0.23 \textrm{s}^{-1}$.
Here the values of $T$ and $B_\bot$ are chosen to guarantee
$Tw_{-9/2,7/2;-9/2,9/2}=1$, \emph{i.e.}, the transverse magnetic
pulse introduced is a $\pi$ pulse. From the data above, we can
neglect the transitions between the states $(9/2,9/2)$ and
$(9/2,7/2)$ compared with that between $(-9/2,9/2)$ and
$(-9/2,7/2)$. In addition, the pulsed transverse magnetic field
can induce transitions from the states as quantum computing basis
to the other excited states, which lead to the decaying of quantum
computing basis. When $T=10^{-8}\textrm{s}$ and
$B_\bot=3.8\textrm{G}$, the transition rates
$w_{-9/2,5/2;-9/2,7/2},w_{9/2,5/2;9/2,7/2}$ and
$w_{7/2,7/2;9/2,7/2}$ are $4.4 \times 10^1\textrm{s}^{-1}, 8.8
\textrm{s}^{-1}$ and $3.6\times 10^2 \textrm{s}^{-1}$
respectively; when $T=10^{-7}\textrm{s}$ and
$B_\bot=0.38\textrm{G}$, transition rates
$w_{-9/2,5/2;-9/2,7/2},w_{9/2,5/2;9/2,7/2}$ and
$w_{7/2,7/2;9/2,7/2}$ are $2.5\times 10^{-2} \textrm{s}^{-1}$,
$4.3\times 10^{-2} \textrm{s}^{-1}$ and $3.6\times 10^{-1}
\textrm{s}^{-1}$ respectively. These transition rates are smaller
several orders than the transition rates $w_{-9/2,7/2;-9/2,9/2}$
and $w_{-9/2,9/2;-9/2,7/2}$, so they are also negligible.

From the above discussion, we know that,  when introducing  the
special frequency transverse magnetic field, only coherent
transition between the spin states $(-9/2,9/2)$ and $(-9/2,7/2)$
is prominent and others are negligible. So we can interpret it as
Rabi oscillation of two level atom. With $|a\rangle$ and
$|b\rangle$ denote the spin states $(-9/2,9/2)$ and $(-9/2,7/2)$
respectively, simplify the Hamiltonian as
\begin{eqnarray}
H_{rabi}(t)&=&\hbar\omega_a|a\rangle\langle a|+\hbar\omega_b|b\rangle\langle b|\nonumber\\
&&-\frac{\hbar}{2}\Omega(e^{i\omega t}|a\rangle\langle
b|+e^{-i\omega t}|b\rangle\langle a|). \label{hrabi}
\end{eqnarray}
where $\omega_a=E^{(0)}_{-9/2,9/2}/\hbar$,
$\omega_b=E^{(0)}_{-9/2,7/2}/\hbar$ and $\Omega=
g\mu_BB_\bot\hbar$ is Rabi frequency. The coherent wave function
of the two states can be written in the form
\begin{equation}
|\psi(t)\rangle=c_a(t)e^{-i\omega_a
t}|a\rangle+c_b(t)e^{-i\omega_b t}|b\rangle. \label{wave}
\end{equation}
From Eqs.(\ref{hrabi}) and(\ref{wave}), we obtain the solution
\begin{eqnarray}
c_a(t)=c_a(0)\cos\left(\frac{\Omega t}{2}\right)+i
c_b(0)\sin\left(\frac{\Omega t}{2}\right),\\
c_b(t)=c_b(0)\cos\left(\frac{\Omega t}{2}\right)+i
c_a(0)\sin\left(\frac{\Omega t}{2}\right),
\end{eqnarray}
where $c_a(0)$ and $c_b(0)$ are the initial values of $c_a$ and
$c_b$ respectively when $t=0$. From this solution, we know that,
if the transverse magnetic field introduced is $\pi$ pulse, i.e.
$\Omega t=\pi$, we realize a NOT gate between the two states.
Simultaneously, other states of computing basis do not vary.

\begin{table}[ht]
\begin{center}
\caption{The comparisons of  the physical quantum states and the
quantum logic states of qubits for a controlled-NOT gate}
\label{tab2}
\begin{tabular}{ccc}\hline\hline
Physical quantum states&\hspace{2mm} Quantum logic states\\
\hline $(9/2,9/2)\longrightarrow (9/2,9/2)$
 &$|00\rangle\longrightarrow|00\rangle$\\
$(9/2,7/2)\longrightarrow
(9/2,7/2)$ &$|01\rangle\longrightarrow|01\rangle$\\
$(-9/2,9/2)\longrightarrow (-9/2,7/2)$
&$|10\rangle\longrightarrow|11\rangle$\\
$(-9/2,7/2)\longrightarrow (-9/2,9/2)$ &$|11\rangle\longrightarrow|10\rangle$\\
\hline\hline
\end{tabular}
\end{center}
\end{table}

Therefore,the pulsed magnetic field gives rise to the state
transitions shown in the left column of  Table \ref{tab2}. We
choose the first $\textrm{Mn}_4$ unit as the control  qubit and
the second one as the target qubit. Here the quantum states
$m_1=9/2$ and $-9/2$ of the first $\textrm{Mn}_4$ unit correspond
to the quantum logic state $|0\rangle$ and $|1\rangle$ of the
control qubit respectively, while the quantum states $m_2=9/2$ and
$7/2$ of the second $\textrm{Mn}_4$ unit correspond to the quantum
logic state $|0\rangle$ and $|1\rangle$ of the target qubit
respectively, as shown in Table \ref{tab2}. In fact, the
transition induced by the pulsed magnetic field correspond to the
transform of the the quantum computing basis as
\begin{equation}
U_{CNOT}=\left(\matrix{1&0&0&0\cr 0&1&0&0\cr 0&0&0&1\cr
0&0&1&0}\right).
\end{equation}
So our scheme has realized a conditional quantum dynamics  in a
dimer of exchange coupled single-molecule magnets,
$[\textrm{Mn}_4]_2$.

In our scheme a key point is that the frequency of the pulsed
transverse magnetic field is chosen as $\omega=\omega_1$ instead
of $\omega=\omega_2$. Seemingly, if we  choose the quantum states
$m_1=-9/2$ and $9/2$ of the first $\textrm{Mn}_4$ unit as the
quantum logic states $|0\rangle$ and $|1\rangle$ of the control
qubit respectively and the frequency of the pulsed transverse
magnetic field as $\omega=\omega_2$, the controlled-NOT gate can
also be realized. However, this is not true. In fact, when
$\omega=\omega_2$, the pulsed transverse field can induce the
transition from the state $(9/2,9/2)$ to the states $(9/2,7/2)$ or
$(7/2,9/2)$ by absorbing a $\sigma^-$ photon, because the states
$(9/2,7/2)$ and $(7/2,9/2)$ are energy degenerate and the energy
gaps $E^{(0)}_{9/2,7/2}-E^{(0)}_{9/2,9/2}$ and
$E^{(0)}_{7/2,9/2}-E^{(0)}_{9/2,9/2}$ are identical with each
other, which are shown in Table \ref{tab1}. Thus, if we choose
$\omega=\omega_2$, the state $(9/2, 9/2)$ will decay into the
state $(7/2,9/2)$, which does not belong to the quantum computing
basis.

{\it The effects of the transverse exchange interactions}-- In the
above discussion, we have not considered the transverse exchange
interactions, \textit{i.e.}, the last term in Eq.(\ref{dimer}),
re-denoted as
\begin{equation}
H_{xy}=\frac{J_{xy}}{2}(S^+_{1}S^-_{2}+S^-_{1}S^+_{2})
\label{hxy}
\end{equation}
 which in fact can induce the decaying of  the quantum computing basis
into other excited  states. In the first order, $H_{xy}$ acts
between the zeroth-order eigenvectors $(m_1,m_2)$ and $(m_1\pm
1,m_2\mp 1)$. The effect of $H_{xy}$ on the tunnelling of the
states is discussed in details in Ref.\cite{hill}. We
perturbatively evaluated the amplitude of the transition  from the
states from $(l,l')$ to $(k,k')$ as,
\begin{eqnarray}
C_{kk';ll'}=\frac{2\pi}{i\hbar}\langle kk'|
H_{xy}|ll'\rangle\delta^{(T)}(\omega_{kk',ll'}) \label{chxy}
\end{eqnarray}
 Here the transitions from the quantum computing basis to other excited states induced by the
$H_{xy}$ are $(9/2,7/2)\rightarrow(7/2,9/2),\;
(-9/2,9/2)\rightarrow(-7/2,7/2)$ and
$(-9/2,7/2)\rightarrow(-7/2,5/2)$, while transition from the state
$(9/2,9/2)$ can not occur since the total spin of the
$\textrm{Mn}_4$ unit is $9/2$. Since the superexchange interaction
of the dimer $[\textrm{Mn}_4]_2$ is nearly isotropic\cite{tiron},
we set the parameter $J_{xy}=0.1K$. When the duration $T$ is
infinite, $\delta^{(T)}(\omega_{kk',ll'})$ in Eq.(\ref{chxy})
becomes Dirac delta function $\delta(\omega_{kk',ll'})$. The
energy conservation  holds when the transitions happen. From Table
\ref{tab1}, only the transition between  spin states $(9/2,7/2)$
and $(7/2,9/2)$ is possible. However, if  $T$ is finite, the
energy conservation does not hold during transition due to
uncertainty principle. Thus, the transition between nondegenerate
states is possible if  $T$ is finite. Because the time of quantum
computing operation is finite, it is necessary to evaluate the
transition rate due to exchange interaction to compare with
transition induced by magnetic field.
 When $T$ is
$10^{-8}\textrm{s}$, we evaluated the transition rates
$w_{7/2,9/2;9/2,7/2}$, $w_{-7/2,7/2;-9/2,9/2}$ and
$w_{-7/2,5/2;-9/2,7/2}$ are $8.9\times 10^{-13}\textrm{s}^{-1}$,
$3.8\times 10^{-21}\textrm{s}^{-1}$ and $1.3\times
10^{-20}\textrm{s}^{-1}$ respectively.  If the duration
$T=10^{-7}\textrm{s}$, then $w_{7/2,9/2;9/2,7/2}=8.9\times
10^{-12}\textrm{s}^{-1}$, $w_{-7/2,7/2;-9/2,9/2}=6.2\times
10^{-22}\textrm{s}^{-1}$ and $w_{-7/2,5/2;-9/2,7/2}=1.1\times
10^{-21}\textrm{s}^{-1}$. The rates of transitions induced by the
transverse exchange interactions are  far smaller  than that of
the transitions between the quantum computing basis, so we can
neglect them and do not consider their effects in our scheme.

{\it Conclusion}-- We have proposed a scheme to realize a
controlled-NOT quantum logic gate in a dimer of exchange coupled
single-molecule magnets, $[\textrm{Mn}_4]_2$. We first neglected
the transverse exchange interactions between the two
$\textrm{Mn}_4$ units and obtained the spin states, and the energy
spectrum. Then, we chosen the ground state and the three low-lying
excited states in a finite longitudinal magnetic field as the
quantum computing basis  and introduced a pulsed transverse
magnetic field with a special frequency, which can induce
transitions between the quantum computing basis so as to realize a
controlled-NOT operation. In our scheme, the magnetic pulse is a
$\pi$ pulse, which leads to the transitions of spin states with
$\Delta m=\pm1$. We have evaluated  the transition rates induced
by the transverse exchange interactions and analyzed the their
effects on decaying of the states. In this paper, we have not
considered the initializing, read-in and read-out of the states,
which are needed to improve in technique for molecular magnets. If
the measure approach of the single molecular magnet is improved,
molecular magnets are promising candidates for quantum computing.

This work is in part supported by NSF of China No.10447125.

\end{document}